\documentclass{PoS}

\PoS{PoS(LAT2005)329}

\title{Gauge invariance of the Abelian dual Meissner effect in pure SU(2) QCD}

\ShortTitle{Gauge invariance of the Abelian dual Meissner effect in pure SU(2) QCD}

\author{\speaker{Toru Sekido}\\
        Institute for Theoretical Physics, Kanazawa University, Kanazawa
        920-1192, Japan\\
        and RIKEN, Radiation Laboratory, Wako 351-0158, Japan\\
        E-mail: \email{toryu@hep.s.kanazawa-u.ac.jp}}

\author{Katsuya Ishiguro, Yoshifumi Nakamura and Tsuneo Suzuki\\
        Institute for Theoretical Physics, Kanazawa University, Kanazawa
        920-1192, Japan \\
        and RIKEN, Radiation Laboratory, Wako 351-0158, Japan}

\abstract{The dual Meissner effect is described and numerically
observed in a gauge-invariant way in lattice Monte-Carlo simulations in pure $SU(2)$ QCD.
The squeezing of the non-Abelian electric field between a pair of static quark and anti-quark occurs
due to the solenoidal current coming from the gauge-invariant monopole-like quantity.
Preliminary results are obtained with respect to the vacuum type of the confinement phase.
The $SU(2)$ QCD vacuum seems near the border between the type 1 and the type 2 dual superconductors.
The theoretical background of this idea is published in another report~\cite{Suzuki:2005lat051}. Here we show numerical results in this note.
}

\FullConference{XXIIIrd International Symposium on Lattice Field Theory\\
		 25-30 July 2005\\
		 Trinity College, Dublin, Ireland}

\newcommand{\beqn}{\begin{eqnarray}}
\newcommand{\eeqn}{\end{eqnarray}}

\begin{document}

\section{Introduction}
In a previous report~\cite{Suzuki:2005lat051},
we have defined a gauge-invariant monopole-like quantity which we call as 'monopole' through a gauge-invariant Abelian-like field strength.
The definition of the Abelian-like field strength $f_{\mu\nu}$ and
'monopole'
$k_{\mu}$ on the lattice 
are the followings:
\begin{eqnarray}
f_{\mu\nu}(s)&=&n^a_{\mu\nu}(s)F^a_{\mu\nu}(s) \label{lattice-fmunu}, \\
k_{\mu}(s)&=&\frac{1}{8\pi}\epsilon_{\mu\nu\alpha\beta}\Delta_{\nu}
f_{\alpha\beta}(s+\hat{\mu}). \label{lattice-monopole1}
\end{eqnarray}

The purpose of this paper is to show the numerical evidence 
that the dual Meissner effect occurs in a gauge-invariant way
due to the gauge-invariant Abelian-like field strength and 'monopoles'.
We do not need any Abelian projection nor any gauge-fixing.

\begin{figure}[b]
\includegraphics[height=8cm]{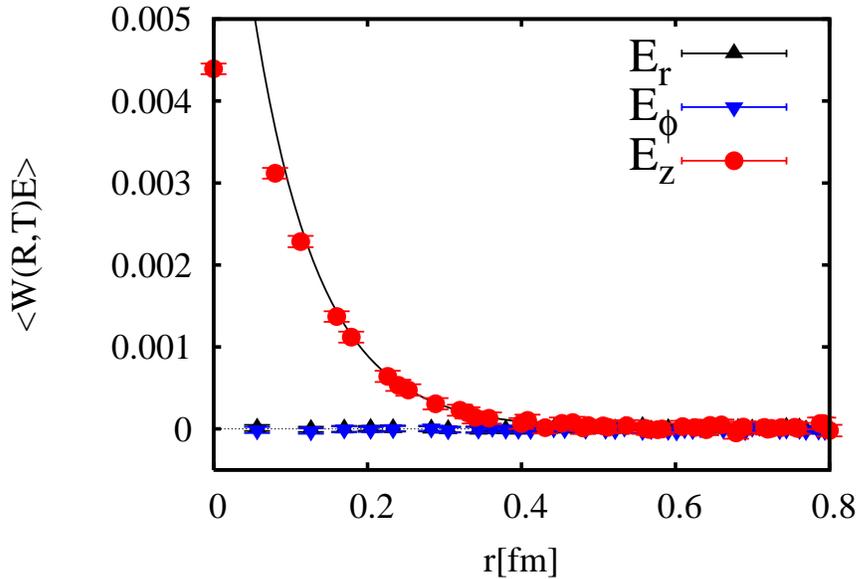}
\caption{\label{fig-E} $\vec{E}$ electric field  profiles. $r$ is a distance perpendicular to the $Q\bar{Q}$ axis and 
 $W(R\times T= 5\times 5)$ is used.}
\end{figure}

\section{Numerical results}

\subsection{Method}

We use RG improved gluonic action~\cite{Iwasaki:1985we,Iwasaki:1983ck},
\begin{eqnarray}
S=\beta\left\{
 c_0\sum \mbox{Tr}\left(plaquette\right)
+c_1\sum \mbox{Tr}\left(rectangular\right)
\right\}.
\end{eqnarray}
The mixing parameters are fixed as $c_0+8c_1=1$ and $c_1=-0.331$.
We adopt the coupling constant $\beta=1.20$ which corresponds to the
lattice distance $a(\beta=1.20)=0.0792(2)$ [fm].
The lattice size is $32^4$.
After 5000 thermalizations, we have taken 2000 thermalized configurations per 100 sweeps for measurements. 
To get a good signal-to-noise ratio, the APE smearing technique~\cite{Albanese:1987ds}
is used for evaluating Wilson loops.

\begin{figure}[htb]
\begin{minipage}{.48\textwidth}
\hspace{0.8cm}
\includegraphics[width=1.0\textwidth]{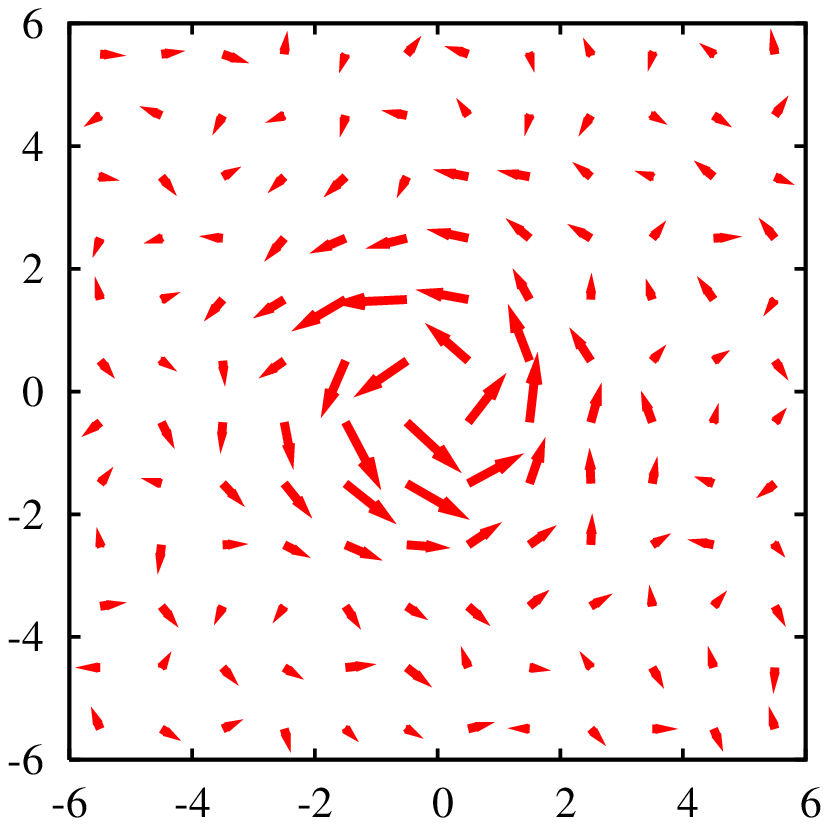}
\caption{\label{fig-veck}'Monopole' current distributions.}
\end{minipage}
\hspace{.04\textwidth}
\begin{minipage}{.48\textwidth}
\includegraphics[width=1.0\textwidth]{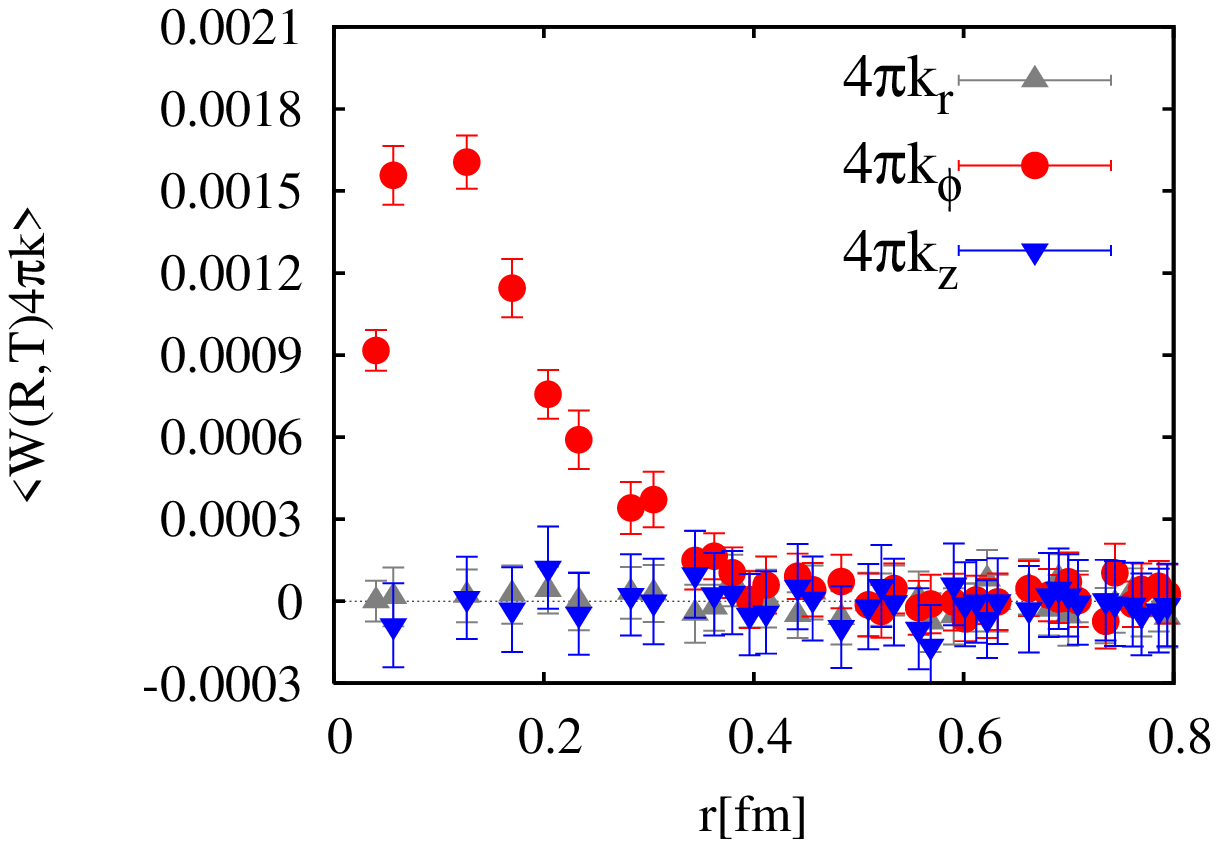}
\caption{\label{fig-k}Components of 'monopoles' around a static quark
 pair.}
\end{minipage}
\end{figure}
\begin{figure}[h]
\begin{minipage}{.48\textwidth}
\includegraphics[width=1.0\textwidth]{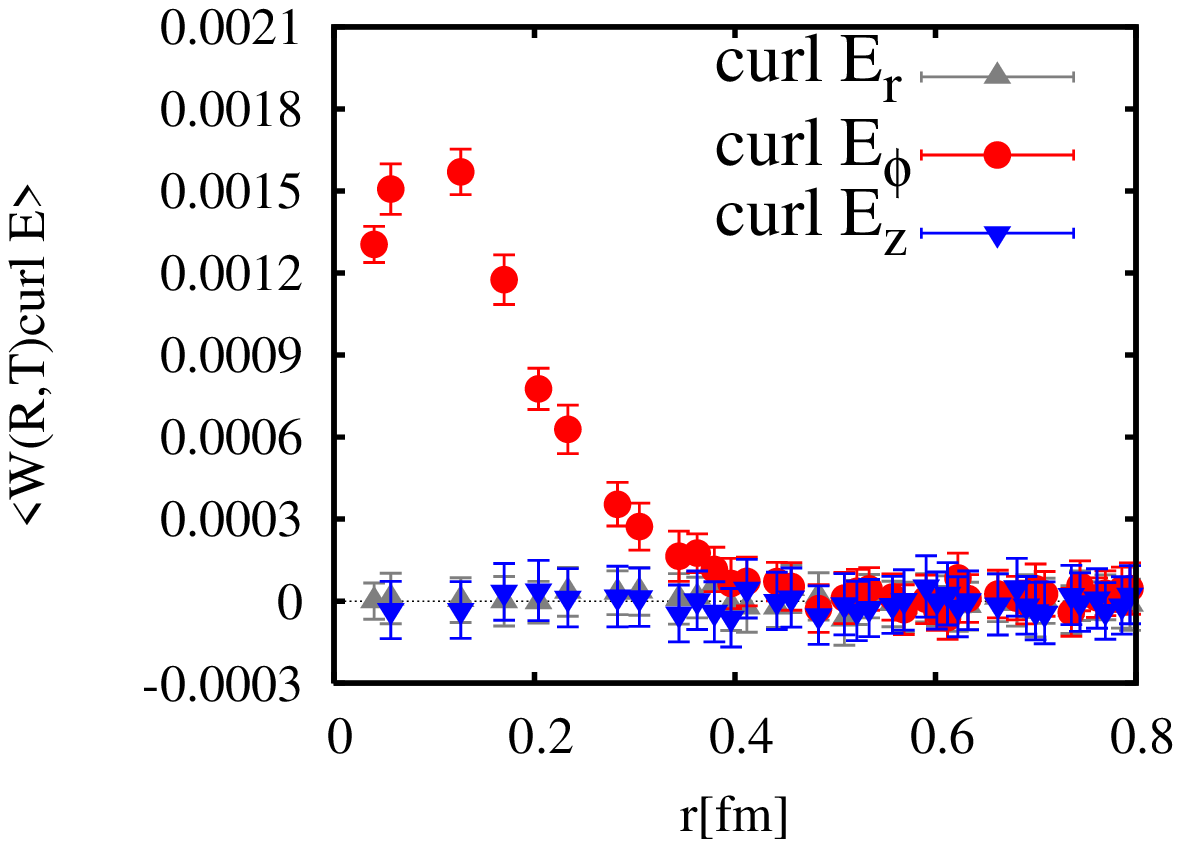}
\caption{\label{fig-rotE}Components of curl$\vec{E}$ around a static quark pair.}
\end{minipage}
\hspace{.04\textwidth}
\begin{minipage}{.48\textwidth}
\includegraphics[width=1.0\textwidth]{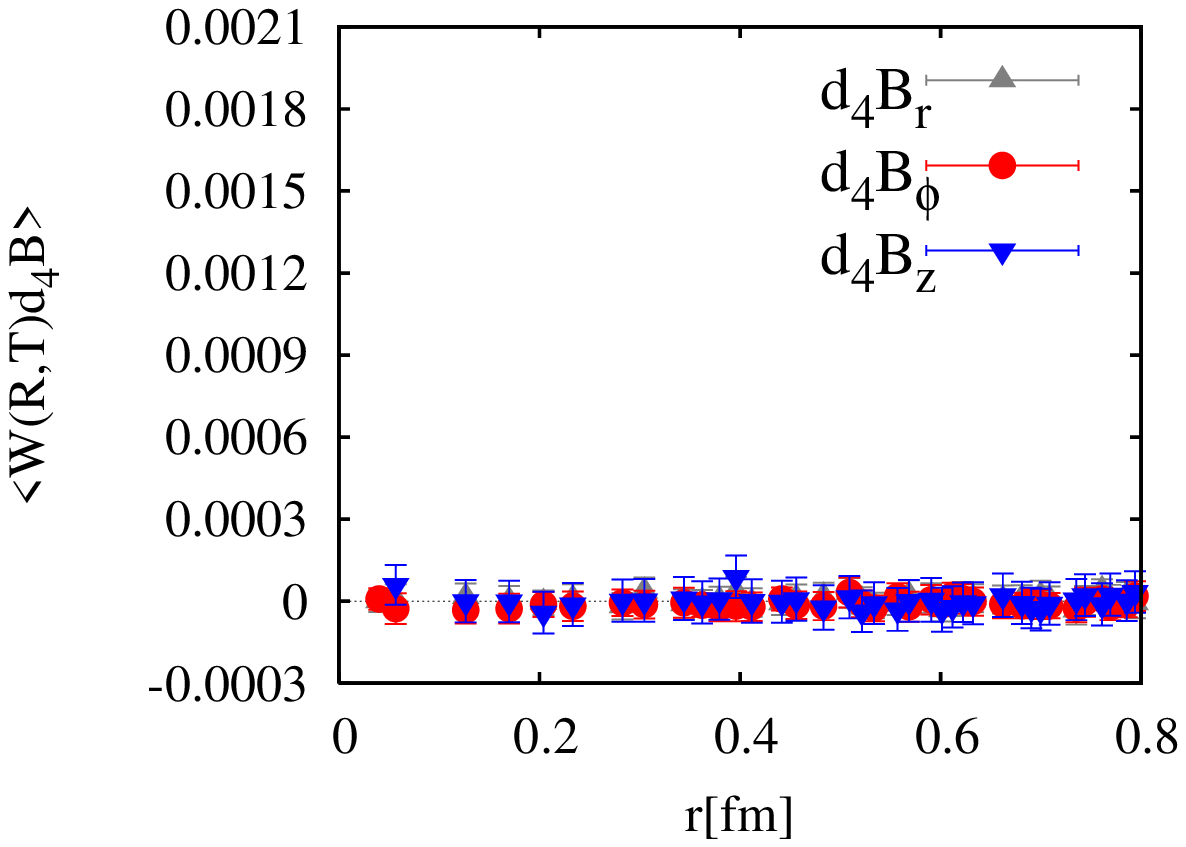}
\caption{\label{fig-dtB}Components of the magnetic displacement current around a static quark pair.}
\end{minipage}
\end{figure}

\subsection{Correlations with Wilson loops}
Let us try to measure, without any gauge fixing, 
electric and magnetic flux distributions by evaluating correlations of
Wilson loops 
and the Abelian-like field strengths located in the perpendicular direction to the Wilson-loop plane.
First we show in Fig.\ref{fig-E}  electric  field profiles around a 
quark pair. Only the $z$-component of the electric field is non-vanishing and squeezed.
The profiles are studied mainly on a perpendicular plane at the midpoint between the 
two quarks. Note that electric fields perpendicular to the $Q\bar{Q}$
axis are found to be negligible. 
The solid line denotes the best exponential fit.
 
\begin{figure}[t]
\includegraphics[height=8.cm]{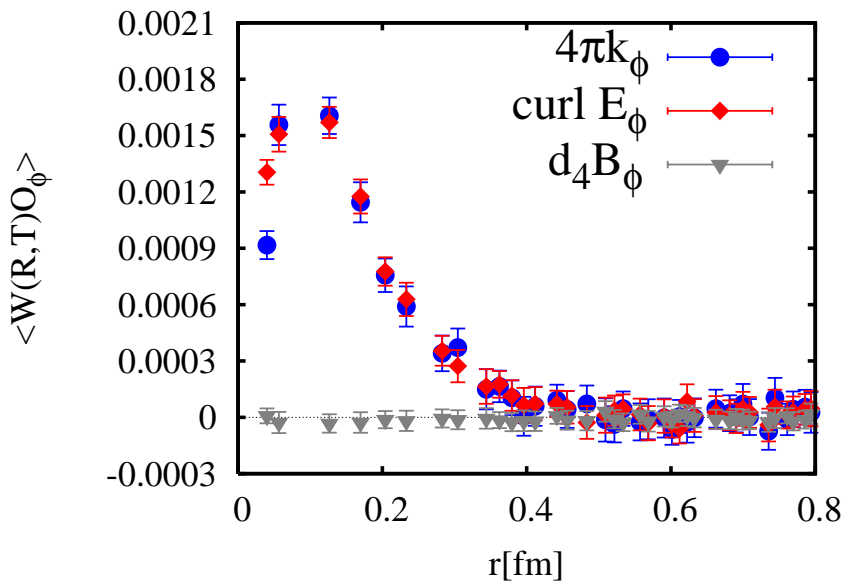}
\caption{\label{fig-kk}The azimuthal component of 'monopoles' and the magnetic displacement current around a static quark pair.}
\end{figure}

\begin{figure}[b]
\begin{minipage}{.48\textwidth}
\includegraphics[width=1.0\textwidth]{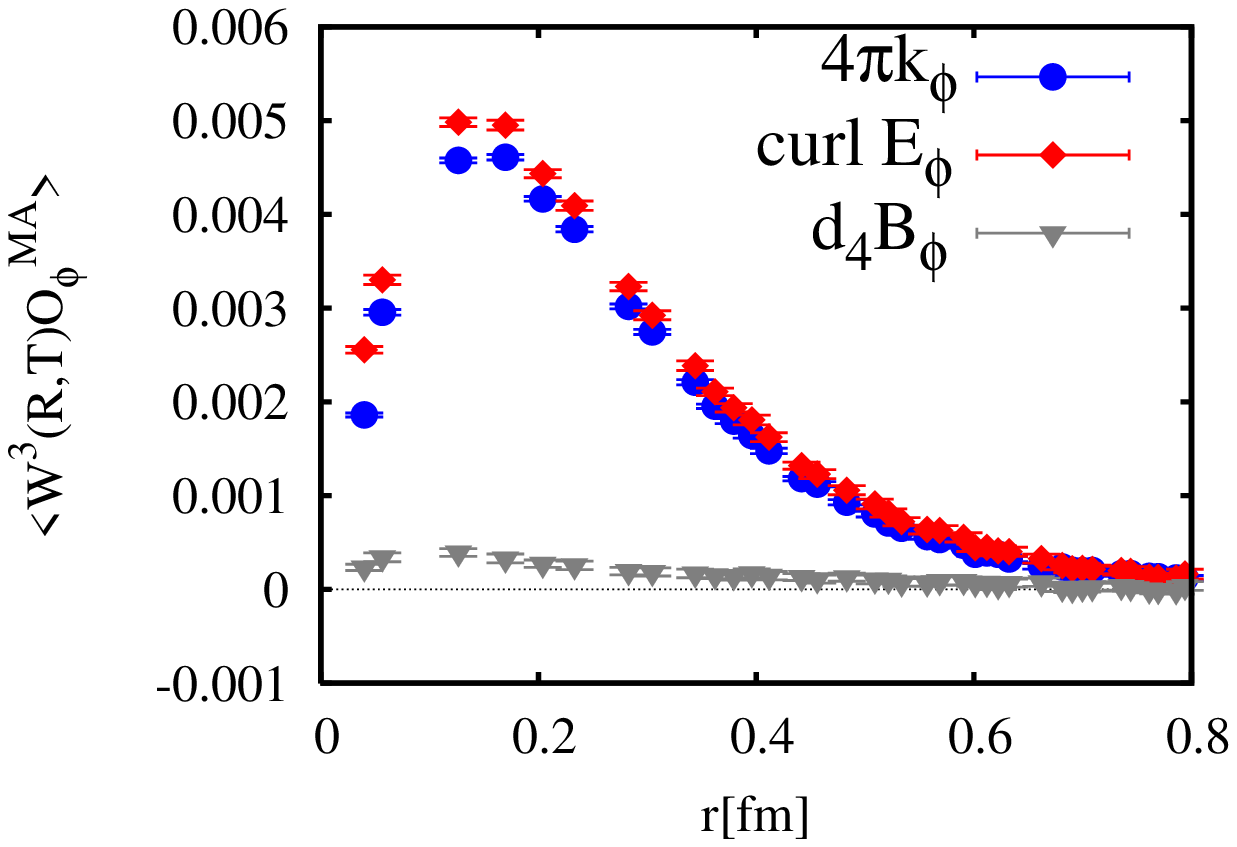}
\caption{\label{fig-kkMA}The azimuthal component of the Abelian monopoles and the magnetic displacement current  around a static quark pair in the MA gauge.}
\end{minipage}
\hspace{.04\textwidth}
\begin{minipage}{.48\textwidth}
\includegraphics[width=1.0\textwidth]{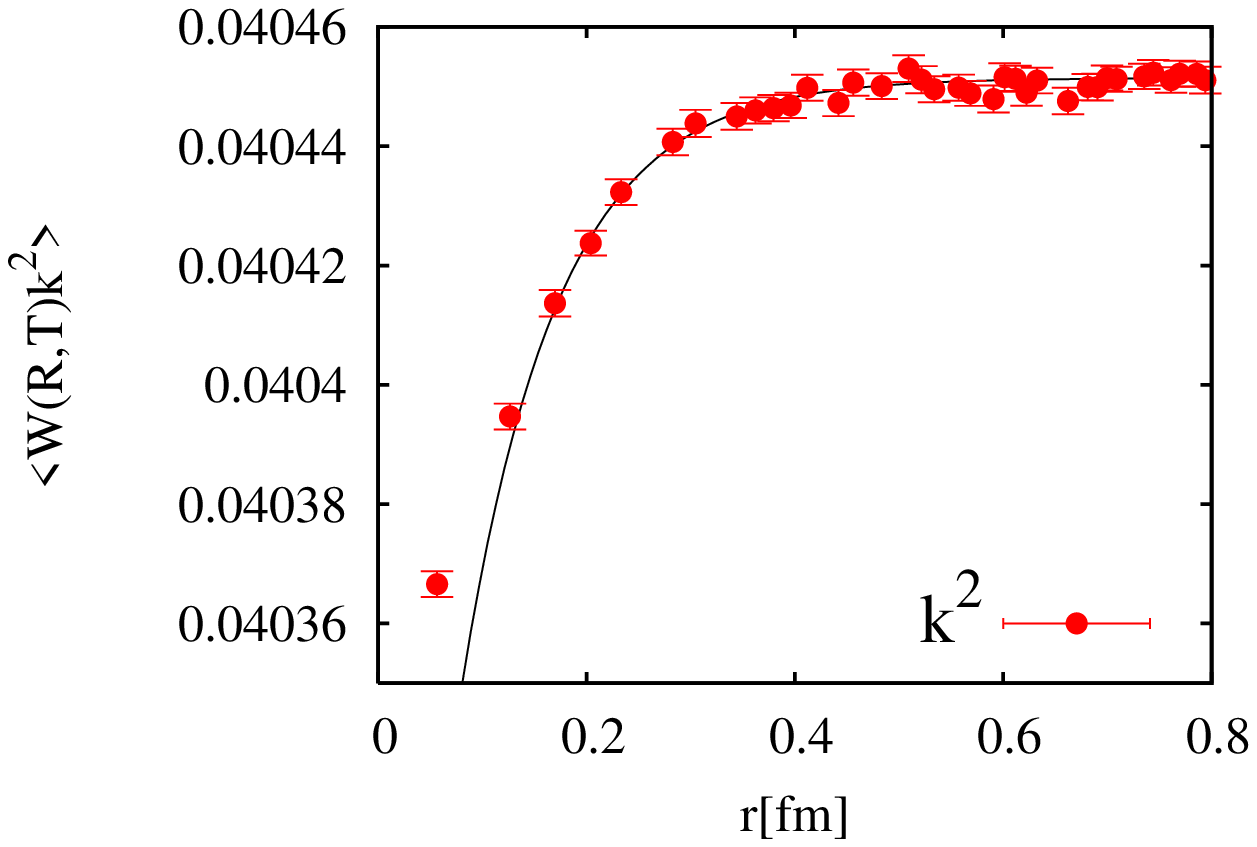}
\caption{\label{fig-k2}The correlation between the Wilson loop and the squared monopole density. $W(R=5,T=5)$ is used.}
\end{minipage}
\end{figure}

Now let us study the violation of the Bianchi identity with respect to
the Abelian-like field strength.
\begin{eqnarray}
\vec{\nabla}\times\vec{E}+\partial_{4}\vec{B}&=&4\pi\vec{k}. \label{BI}
\end{eqnarray}
The Coulombic electric field coming from the static source is written in
the lowest perturbation theory in terms of the gradient of a scalar
potential. 
 Hence it does not contribute to the curl of the  electric field nor to the  magnetic field in the above  Bianchi identity  Eq.(\ref{BI}). 
The dual Meissner effect  says that the squeezing of the electric flux
occurs due to cancellation of the Coulombic electric fields and those
from  solenoidal magnetic currents. 
It is very interesting to see from Fig.\ref{fig-veck} that in this gauge-invariant case, the gauge-invariant 'monopole' Eq.(\ref{lattice-monopole1}) plays the role of the solenoidal current. This is qualitatively similar to  the monopole behaviors in the MA gauge
 ~\cite{Singh:1993jj,Bali:1996dm,Koma:2003gq,Koma:2003hv}. 
 
 Let us see also the $r$ dependence of the 'monopole' distribution shown
 in  Fig.\ref{fig-k}. All $r$ and $z$ components of each term in
 Eq.(\ref{BI}) are almost vanishing  consistently with
 Fig.\ref{fig-veck}. 
The all components of Eq.(\ref{BI}) is plotted in Fig.\ref{fig-rotE} and Fig.\ref{fig-dtB}.
The magnetic displacement current $\partial_4\vec{B}$ are found to be
 negligible numerically as similarly as in the MA
 gauge~\cite{Singh:1993jj,Bali:1996dm,Cea:1995zt}. We show in 
Fig.\ref{fig-kk} 
and Fig.\ref{fig-kkMA}
azimuthal components of  all three terms of Eq.(\ref{BI})
in this gauge-invariant case and in the MA gauge case\footnote{In the MA case, we use only the third component $W^3$ of the non-Abelian Wilson loop as a source.}.  
 It is interesting that the peak positions of gauge-invariant $k_{\phi}$
and $k_{\phi}^{MA}$ in the MA gauge look similar around $0.15$ [fm], although the height and 
the shapes seem  different. 

\subsection{The type of the vacuum}
Let us next try to fix the type of the vacuum of pure $SU(2)$ QCD.
The penetration length $\lambda$ is determined by making an exponential fit to the electric field flux for large $r$ regions.
 The best fitting curve is also plotted in Fig.~\ref{fig-E} from which we fix the penetration length $\lambda $
\footnote{We have fixed both lengths using a simple exponential function expected in the long-range regions. For details, see Ref.~\cite{Chernodub:2005gz}.}.
Next  we derive the coherence length $\xi$. In 
Ref.~\cite{Chernodub:2005gz},
we have shown that the coherence length can be fixed by a measurement of the squared monopole density  around the $Q\bar{Q}$ pair. The same situation is expected in this gauge-invariant case. 
The correlation between the Wilson loop and the squared 'monopole' density  is plotted in
Fig.~\ref{fig-k2}. From the exponential fit, we may fix the coherence
length. 
We get (in Table~\ref{tbl:best:fits})
\begin{eqnarray}
\lambda=0.085(5) [fm], \\
\xi/\sqrt{2}=0.10(1) [fm].
\end{eqnarray}
Although  the $R$ and $T$ dependences of both lengths are not studied yet, the value of the coherence length looks almost the same as that of the penetration length within the error bars.  Hence if the same situations will continue for larger $R$  in the confining string region, the type of the vacuum is fixed to be near the border between the type 1 and the type 2. This is consistent with the result of our previous paper~\cite{Chernodub:2005gz} and the  result~\cite{Haymaker:2005py} obtained in the MA gauge.  
It should be stressed that the present result is obtained in  a gauge-invariant framework.  
\begin{table*}[htb]
\begin{center}
\begin{tabular}{|c|c|c|c|c|c|}
\hline
Quantity                          &  Fig.        & $c_0$      & $c_1$ [fm]  & $c_2$        & $\chi^2/d.o.f.$ \\
\hline
$\langle W E_z \rangle          $ & \ref{fig-E}  & $0.009(2)$ & $0.085(5)$ & $0$          & $4.2$    \\
$\langle W k^{2} \rangle $ & \ref{fig-k2} & $0.010(2)$ & $0.10(1)$  & $1.59534(1)$ & $4.1$   \\
\hline
\end{tabular}
\caption{\label{tbl:best:fits} The best fit parameters corresponding to
 the fits of various quantities by a function $f(r)=c_0
 \exp(-r/c1)+c_2$.}
\end{center}
\end{table*}

\section{Conclusions}
Monte-Carlo simulations of quenched $SU(2)$ QCD are performed.
It has been found that the squeezing of the non-Abelian electric field $\sqrt{(E^a_i)^2}$
occurs and the solenoidal current from the gauge-invariant 'monopoles' is responsible for the flux squeezing.
The magnetic displacement current observed previously in Landau gauge~\cite{Suzuki:2004dw} has been found to be negligible.
Preliminary results have been obtained with respect to the vacuum type of the confinement phase.
The $SU(2)$ QCD vacuum seems near the border between the type 1 and the type 2 dual superconductors.
This is consistent with the result of our previous paper~\cite{Chernodub:2005gz} 
and the result~\cite{Haymaker:2005py} obtained in the MA gauge.  
Present numerical results are not perfect, since the continuum limit,
the infinite-volume limit and the real $SU(3)$ case are not studied yet. 
Nevertheless the results obtained here are very interesting,
since they show for the first time the flux squeezing of non-Abelian electric fields
is working in a gauge-invariant way due to the dual Meissner effect without performing any Abelian projection.

\begin{acknowledgments}
The numerical simulations of this work were done using RSCC computer clusters in 
RIKEN. The authors would like to thank RIKEN for their support of computer facilities. 
T.S. is supported by JSPS Grant-in-Aid for Scientific Research on Priority Areas 13135210 and (B) 15340073.
\end{acknowledgments}


\begin{thebibliography}{10}

\bibitem{Suzuki:2005lat051}
T.~Suzuki, K.~Ishiguro, Y.~Nakamura, and T.~Sekido,
\newblock {\em Gauge invariant 'monopoles' and color confinement mechanism},
\newblock in {\em Proceedings of Lattice 2005 conference}, 2005.

\bibitem{Iwasaki:1985we}
Y.~Iwasaki,
\newblock {\em Renormalization group analysis of lattice theories and improved lattice action: two-dimensional nonlinear O(N) sigma model},
\newblock {\em Nucl. Phys.} {\bf B258}, 141 (1985).

\bibitem{Iwasaki:1983ck}
Y.~Iwasaki,
\newblock {\em Renormalization group analysis of lattice theories and
	improved lattice action: four-dimensional non-Abelian SU(N) gauge model},
\newblock 1983.

\bibitem{Albanese:1987ds}
APE, M.~Albanese {\em et~al.},
\newblock {\em Glueball masses and string tension in lattice QCD},
\newblock {\em Phys. Lett.} {\bf B192}, 163 (1987).

\bibitem{Singh:1993jj}
V.~Singh, D.~A. Browne, and R.~W. Haymaker,
\newblock {\em Structure of Abrikosov vortices in SU(2) lattice gauge theory},
\newblock {\em Phys. Lett.} {\bf B306}, 115 (1993),
\newblock [{\tt hep-lat/9301004}].

\bibitem{Bali:1996dm}
G.~S. Bali, V.~Bornyakov, M.~Muller-Preussker, and K.~Schilling,
\newblock {\em Dual superconductor scenario of confinement: a systematic study of gribov copy effects},
\newblock {\em Phys. Rev.} {\bf D54}, 2863 (1996),
\newblock [{\tt hep-lat/9603012}].

\bibitem{Koma:2003gq}
Y.~Koma, M.~Koma, E.-M. Ilgenfritz, T.~Suzuki, and M.~I. Polikarpov,
\newblock {\em Duality of gauge field singularities and the structure of the flux tube in Abelian-projected SU(2) gauge theory and the dual Abelian Higgs model},
\newblock {\em Phys. Rev.} {\bf D68}, 094018 (2003),
\newblock [{\tt hep-lat/0302006}].

\bibitem{Koma:2003hv}
Y.~Koma, M.~Koma, E.-M. Ilgenfritz, and T.~Suzuki,
\newblock {\em A detailed study of the Abelian-projected SU(2) flux tube and its dual Ginzburg-Landau analysis},
\newblock {\em Phys. Rev.} {\bf D68}, 114504 (2003),
\newblock [{\tt hep-lat/0308008}].

\bibitem{Cea:1995zt}
P.~Cea and L.~Cosmai,
\newblock {\em dual superconductivity in the SU(2) pure gauge vacuum: a lattice study},
\newblock {\em Phys. Rev.} {\bf D52}, 5152 (1995),
\newblock [{\tt hep-lat/9504008}].

\bibitem{Chernodub:2005gz}
M.~N. Chernodub {\em et~al.},
\newblock {\em Vacuum type of SU(2) gluodynamics in maximally Abelian and Landau gauges},
\newblock 2005,
\newblock [{\tt hep-lat/0508004}].

\bibitem{Haymaker:2005py}
R.~W. Haymaker and T.~Matsuki,
\newblock {\em Consistent definitions of flux and the dual superconductivity parameters in SU(2) lattice gauge theory},
\newblock 2005,
\newblock [{\tt hep-lat/0505019}].

\bibitem{Suzuki:2004dw}
T.~Suzuki, K.~Ishiguro, Y.~Mori, and T.~Sekido,
\newblock {\em The dual Meissner effect and Abelian magnetic displacement currents},
\newblock {\em Phys. Rev. Lett.} {\bf 94}, 132001 (2005),
\newblock [{\tt hep-lat/0410001}].

\end{thebibliography}

\end{document}